# Relay Assisted F/TDMA Ad Hoc Networks: Node Classification, Power Allocation and Relaying Strategies*

Semih Serbetli[†]          Aylin Yener[‡]

November 29, 2006

## Abstract

This paper considers the design of relay assisted F/TDMA ad hoc networks with multiple relay nodes each of which assists the transmission of a predefined subset of source nodes to their respective destinations. Considering the sum capacity as the performance metric, we solve the problem of optimally allocating the total power of each relay node between the transmissions it is assisting. We consider four different relay transmission strategies, namely regenerative decode-and-forward (RDF), nonregenerative decode-and-forward (NDF), amplify-and-forward (AF) and compress-and-forward (CF). We first obtain the optimum power allocation policies for the relay nodes that employ a uniform relaying strategy for all nodes. We show that the optimum power allocation for the RDF and NDF cases are modified water-filling solutions. We observe that for a given relay transmit power, NDF always outperforms RDF whereas CF always provides higher sum capacity than AF. When CF and NDF are compared, it is observed that either of CF or NDF may outperform the other in different scenarios. This observation suggests that the sum capacity can be further improved by having each relay adopt its relaying strategy in helping different source nodes. We investigate this problem next and determine the optimum power allocation and relaying strategy for each source node that relay nodes assist. We observe that optimum power allocation for relay nodes with hybrid relaying strategies provides higher sum capacity than pure RDF, NDF, AF or CF relaying strategies.

[†]Semih Serbetli is with Philips Research Laboratories, Eindhoven, The Netherlands (e-mail: semih.serbetli@philips.com).

[‡]Aylin Yener is with Wireless Communications and Networking Laboratory (WCAN), Department of Electrical Engineering, The Pennsylvania State University, University Park, PA 16802 USA. She is the Corresponding Author (e-mail: yener@ee.psu.edu).

*This paper was presented in part in Conference on Information Sciences and Systems, CISS'05, Baltimore, MD, March 2005, International Conference on Wireless Networks, Communications, and Mobile Computing 2005, Maui, Hawaii, June 2005, and International Conference on Communications, ICC'06, Istanbul, June 2006. This research is supported in part by NSF grants CCF (CAREER)-0237727 and CNS-0508114.

# I. INTRODUCTION

Wireless systems are continuing to evolve towards networks consisting of nodes that communicate without the need for infrastructure [1], [2]. Yet, serious design challenges exist for wireless ad hoc networks in combating the impairments of the wireless channel. Relay assisted communications, where intermediate nodes help forward traffic from source nodes to their destinations, exploits spatial diversity without needing to deploy physical antenna arrays [3]–[11]. Relay assistance also mitigates the effects of path loss, and provides the source nodes with extended battery life [7]–[9], and even help in extending coverage [9], [12], [13]. An intense research effort is underway to better understand the value of relay assisted schemes in rendering the deployment of truly ad hoc wireless networks.

Results related to the capacity of the full duplex relay channel go back to [3], [4]. Since it is difficult to have the nodes transmit and receive simultaneously in the same frequency, much of the recent research effort is towards investigating orthogonal relay transmission schemes where the source and relay nodes transmit in orthogonal channels [5], [7]–[10]. Recently, reference [6] showed that the uplink capacity of two-user systems can be increased by using cooperation, where each user also acts as a relay for the other.

In wireless networks, transmission power of the nodes is limited. Hence, power efficiency is a critical concern when designing relay transmission strategies. It has been shown that significant performance improvement can be achieved by the optimum power allocation for various relay assisted communications systems that consider a *single source destination pair* [7]–[9], [11].

Relay assisted transmission is expected to improve the performance of multiuser systems as well [14]–[17]. Such networks, henceforth referred to as *multiuser relay networks* are ones where each relay node would serve multiple users, and the total transmission power budget for each relay node would be limited. When this is the case, each user's transmission should be relayed with a fraction of the power from its corresponding relay node. In such a scenario, the total relay power should be allocated between the transmissions of information from the sources that relay over this node, in order to obtain the best performance.

In this paper, we consider a wireless ad hoc network with *multiple source-destination pairs*, and relay nodes each of which assists multiple sources. We focus on networks that employ multiaccess techniques with orthogonal transmissions, i.e., F/TDMA. We consider a variety of relay transmission schemes, namely regenerative decode-and-forward (RDF), nonregenerative decode-and-forward (NDF), amplify-and-forward (AF) and compress-and-forward (CF). Considering the sum capacity as the performance metric, we first address the optimum power allocation problem at the relay nodes when all users are assisted via the same relay transmission strategy. We show that the optimal power allocation for RDF and NDF are modified water-filling solutions with a base level and a ceiling, and it classifies the sources into three groups depending on the quality of the source-to-destination, source-to-relay and relay-to-destination links, i.e., *high potential source nodes, low potential source nodes and nonrelayed source nodes*. Next, to obtain higher sum capacities, we investigate the optimum power allocation problem jointly with relaying strategy selection, where the relay node can also choose the relaying strategies for the source nodes it will be assisting, and propose a low complexity near-optimum relaying strategy selection algorithm. We observe that adopting the signaling strategy for each transmission the relay node assists in conjunction with the power level provides significant performance improvement.

## II. System Model and Problem Formulation

We consider a relay assisted F/TDMA ad hoc network with $K$ source nodes (users) and $L$ relay nodes as in Figure 1. We assume that each source node intends to transmit its signal to a destination node and has a pre-assigned relay node that will assist its transmission. The data transmission of each source node occurs in *two* pre-assigned channels that can be either different time slots or different frequencies. The source node broadcasts its signal in the first channel, and the preassigned relay node transmits this source node's information in the second channel. All channels of all source nodes and relay nodes are distinct and nonoverlapping. The signal received by the destination in the $i$th source node's first channel is

$$y_{di1} = \sqrt{P_{si}}\beta_i x_{si} + n_{di1} \qquad (1)$$

where $x_{si}$ is the symbol transmitted by source node $i$, $P_{si}$ is the transmit power of source node $i$ and $\beta_i$ denotes the normalized channel gain from source node $i$ to the destination with $n_{di1}$ as the zero mean AWGN with unit variance. Similarly, the received signal at the relay node $k$ to which source node $i$ is assigned, is

$$y_{ri} = \sqrt{P_{si}}\alpha_i x_{si} + n_{ri} \qquad (2)$$

where $\alpha_i$ is the normalized channel gain from source node $i$ to the assigned relay node $k$, and $n_{ri}$ is the zero mean AWGN with unit variance at the dedicated relay node of source $i$, i.e., $k$. In the second channel of the $i$th source node, the $k$th relay node transmits $x_{ri}$, and the corresponding received signal at the destination is

$$y_{di2} = \sqrt{P_{ri}}\gamma_i x_{ri} + n_{di2} \tag{3}$$

where $x_{ri}$, $P_{ri}$ and $\gamma_i$ denote the signal transmitted for source node $i$ from the $k$th relay node, the transmit power of the $k$th relay node dedicated to source node $i$ and the normalized channel gain from the $k$th relay node to the destination of the $i$th source node with a zero mean and unit variance AWGN $n_{di2}$, respectively. Note that the relay node should transmit after the source due to causality constraints, and this constraint results in loss of one time slot when the channels represent different frequencies. We assume that each relay node has a total power constraint $\sum_{i \in A_k} P_{ri} \le P_{Rk,total}$ where $A_k$ denotes the set of source nodes that relay their information through node $k$. It is also assumed that the relay nodes have the channel state information of the source-to-destination, source-to-relay and relay-to-destination links of all source nodes they serve.

We consider four different relay transmission schemes at the relay nodes, and address the optimum power allocation in each case individually as well as the optimum power allocation with hybrid relaying.

- **Regenerative Decode-and-Forward (RDF)**: When the transmission from the source node is received reliably at the relay node, the relay node decodes the signal, re-encodes it with the same codebook used in the original source node's transmission and transmits the signal in the second channel of the source node [5], [8], [9].

- **Nonregenerative Decode-and-Forward (NDF)**: Similar to RDF, the relay node decodes the received signal, but re-encodes it with a codebook generated independently from that of the source node and transmits it in the second channel of the source node [7].

- **Amplify-and-Forward (AF)**: Perfect decodability at the relay node is not required. The relay node simply forwards the signal in the second channel of the source node by amplifying the received signal at the relay node [5], [9].

- **Compress-and-Forward (CF)**: Similar to AF, the relay node is not required to decode the source's signal perfectly. The relay node compresses the received signal by using Wyner-Ziv lossy source coding, and forwards it in the second channel of the source node [16], [17].

In this work, we aim to optimally distribute the power of each relay node between the source nodes' transmissions to be relayed by that node. Our goal is to maximize the sum capacity of the system. Clearly, the individual capacities of the source nodes are a function of the relay transmission scheme used.

The optimum power allocation problem at the relay nodes is posed as

$$\max_{\{P_{ri}\}_{i=1,\cdots,K}} C_{sum} = \sum_{i=1}^{K} C_{i,*} \tag{4}$$

$$\text{s.t.} \sum_{i \in A_k} P_{ri} \leq P_{Rk,total}; \quad P_{ri} \geq 0 \quad \forall i, k \tag{5}$$

where $C_{i,*}$ is the individual capacity of source node $i$ and $*$ can be replaced with RDF, NDF, AF or CF according to the relay transmission scheme chosen.

In the sequel, we will first assume that all source nodes are assisted by the same relay transmission scheme, and address the optimal power allocation problem for each case. Next, to obtain higher sum capacities, we will remove this assumption and investigate the optimum power allocation problem for hybrid relay nodes where the relay node can assist different source nodes employing different relaying strategies[1]. We note that the F/TDMA architecture allows us to focus on the sum capacity optimization problem at each relay node individually since the power allocation at a relay node does not affect the optimization problem on the others. By optimizing the power allocation at each relay node individually, we optimize the power allocation problem of all relay nodes, since (4) is simply the sum of all sum rates obtained by the assistance of all relay nodes.

## III. Decode and Forward Type Relaying

For both RDF and NDF, the designated relay node must reliably decode the signal. Thus, the individual capacity of a relay assisted source node cannot exceed the capacity of the source node to relay link. This constraint leads to several important observations in terms of optimum power allocation. When the direct link, $\beta_i^2$, is better than the relay link, $\alpha_i^2$ for source node $i$, the minimum of the capacity upper bounds of the direct link and the source node to relay link is the latter. In this case, the capacity of the direct transmission is higher than that of the relay assisted transmission. Since by employing direct transmission for source node $i$, the individual capacity of source node $i$ is maximized, *and* the relay has the potential to improve the sum capacity by investing its power in assisting the remaining source nodes, the relay power allocated to source node $i$ should be

$$P_{ri} = 0 \quad \text{if} \quad \alpha_i^2 \leq \beta_i^2, \quad \forall i = 1, ..., K \tag{6}$$

---

[1]The relay nodes are assumed to be agile radios.

For clarity of exposition, we denote the set of source nodes that are served by the $k$th relay node, and have $\alpha_i^2 > \beta_i^2$ as $\hat{A}_k$ in the sequel. In addition, observe that the maximum individual capacity of source node $i$ is upper bounded by

$$C_{i,RDF} \leq C_{i,NDF} \leq C_{upperDF} = \frac{1}{2K} \log(1 + P_{si}\alpha_i^2), \forall i \tag{7}$$

due to the decodability constraint at the relay. Thus, allocating more power of the relay node for the transmission of a source node beyond a threshold will not increase the individual capacity of the source node. These constraints should be taken into account for the power allocation problem in DF relay nodes.

In the case of RDF relay transmission, the individual capacity of source node $i$ is [5]

$$C_{i,RDF} = min(\frac{1}{2K} \log(1 + P_{si}\beta_i^2 + P_{ri}\gamma_i^2), C_{upperDF}) \tag{8}$$

Similarly, for the case of NDF relay transmission, the individual capacity of source node $i$ can be expressed as [7]

$$C_{i,NDF} = min(\frac{1}{2K} \log(1 + P_{si}\beta_i^2) + \frac{1}{2K} \log(1 + P_{ri}\gamma_i^2), C_{upperDF}) \tag{9}$$

In the sequel, we will use the following classification of source nodes for RDF and NDF networks:

<u>Definition 1</u>: *High potential source nodes (HPU)*: This is the set of source nodes that are allocated nonzero power at their pre-assigned relay node and yet do not achieve the individual capacity upper bound (7). In other words, these are the nodes whose individual capacities would be further increased, if more total power were available at the relay.

<u>Definition 2</u>: *Low potential source nodes (LPU)*: This is the set of source nodes that achieve the maximum individual capacities indicated by (7), by the help of the relay node. For these source nodes, even if more total relay power were available, the individual capacities would not increase.

<u>Definition 3</u>: *Nonrelayed source nodes (NU)*: This is the set of source nodes that are not assisted by the relay node. The source nodes in this set have either high quality direct links, or low quality relay to destination links.

### A. Regenerative Decode and Forward

We are now ready to state our results for RDF relay networks.

**Theorem 1**: The optimal power allocation for RDF relay networks results in three source node sets, namely *high potential source nodes, low potential source nodes*, and *nonrelayed source nodes* for each relay node.

1) The optimum relay power dedicated to <u>high potential</u> source node $i$, and the achieved individual capacity of source node $i$, are

$$P_{ri} = (\frac{1}{\mu_{k,RDF}} - \frac{1 + P_{si}\beta_i^2}{\gamma_i^2})^+; \qquad C_{i,RDF} = \frac{1}{2K}\log(\gamma_i^2/\mu_{k,RDF}) \tag{10}$$

respectively, where $(.)^+ = max(.,0)$ and $\mu_{k,RDF}$ is the water level for the $k$th RDF relay node that satisfies $\sum_{i \in A_k} P_{ri} = P_{Rk,total}$.

2) The optimum relay power dedicated to <u>low potential</u> source node $i$, and the achieved individual capacity of source node $i$, are

$$P_{ri} = \frac{P_{si}(\alpha_i^2 - \beta_i^2)}{\gamma_i^2}; \qquad C_{i,RDF} = \frac{1}{2K}\log(1 + P_{si}\alpha_i^2) \tag{11}$$

3) The <u>nonrelayed</u> source nodes set involves the source nodes that either have better direct links than the source to relay links, i.e., $\alpha_i^2 \leq \beta_i^2$, or high quality direct links or low quality relay to destination links, i.e., $\frac{1 + P_{si}\beta_i^2}{\gamma_i^2} \geq \frac{1}{\mu_{k,RDF}}$.

*Proof:* Using the fact that $P_{ri} = 0$ for the source nodes that have $\beta_i^2 \geq \alpha_i^2$, the optimization problem at the $k$th relay node can be expressed as

$$\max_{\{P_{ri}\}_{i \in \hat{A}_k}} \sum_{i \in \hat{A}_k} \frac{1}{2K}\log(1 + P_{si}\beta_i^2 + P_{ri}\gamma_i^2) \tag{12}$$

$$\text{s.t.} \sum_{i \in \hat{A}_k} P_{ri} \leq P_{Rk,total}; \quad P_{ri} \geq 0, \quad \forall i \tag{13}$$

$$\frac{1}{2K}\log(1 + P_{si}\beta_i^2 + P_{ri}\gamma_i^2) \leq \frac{1}{2K}\log(1 + P_{si}\alpha_i^2), \quad \forall i \tag{14}$$

Constraint (14) is simply an upper bound for $\{P_{ri}\}$, and we thus have

$$0 \le P_{ri} \le \frac{P_{si}(\alpha_i^2 - \beta_i^2)}{\gamma_i^2} \tag{15}$$

Thus, the Lagrangian, $L(\{P_{ri}\}, \hat{\mu}_{k,RDF}, \{\rho_{i,RDF}\})$, is

$$\sum_{i \in \hat{A}_k} \frac{1}{2K} \log(1 + P_{si}\beta_i^2 + P_{ri}\gamma_i^2) + \hat{\mu}_{k,RDF}(P_{Rk,total} - \sum_{i \in \hat{A}_k} P_{ri}) + \sum_{i \in \hat{A}_k} \rho_{i,RDF}(\frac{P_{si}(\alpha_i^2 - \beta_i^2)}{\gamma_i^2} - P_{ri})$$

where $\hat{\mu}_{k,RDF}$ and $\rho_{i,RDF}$ are the Lagrange multipliers associated with the total transmit power constraint of the relay node $k$, and the upper bound for the relay power for source node $i$, respectively. The cost function is a concave function and the $\{P_{ri}\}$ set is a convex set. Thus, simply using the KKT conditions, we arrive at the optimum relay power for source node $i$ as

$$P_{ri} = min((\frac{1}{\mu_{k,RDF}} - \frac{1 + P_{si}\beta_i^2}{\gamma_i^2})^+, \frac{P_{si}(\alpha_i^2 - \beta_i^2)}{\gamma_i^2}) \tag{16}$$

where $\mu_{k,RDF} = 2K\hat{\mu}_{k,RDF}$. The source nodes for which the upper bounds in (14) are inactive, and $P_{ri} = (\frac{1}{\mu_{k,RDF}} - \frac{1 + P_{si}\beta_i^2}{\gamma_i^2}) > 0$, form the high potential source nodes set. When the upper bound is active, $P_{ri} = \frac{P_{si}(\alpha_i^2 - \beta_i^2)}{\gamma_i^2}$, and the corresponding source nodes are the low potential source nodes. Finally, the source nodes with $(\frac{1}{\mu_{k,RDF}} - \frac{1 + P_{si}\beta_i^2}{\gamma_i^2}) \le 0$, or $\alpha_i^2 \le \beta_i^2$ form the set of nonrelayed source nodes. □

**Remark** 1: The optimum power allocation for RDF networks is a modified water-filling solution where each source node has both a base and an upper water level. The base level, $\frac{1 + P_{si}\beta_i^2}{\gamma_i^2}$, is due to the direct link and the channel gain of the relay node to the destination for each source node, whereas the upper level, $\frac{P_{si}(\alpha_i^2 - \beta_i^2)}{\gamma_i^2} + \frac{1 + P_{si}\beta_i^2}{\gamma_i^2}$, is due to the decodability constraints of the RDF relay nodes. Such a power allocation scheme is demonstrated in Figure 2(a) with five source nodes and one relay. In this example, source nodes 1 and 2 are the low potential source nodes for which the relay node allocates enough power for each source node to achieve their maximum individual capacities. Source nodes 3 and 4 are high potential source nodes since their individual capacities can still be improved by increasing the relay power. Source node 5 is a nonrelayed source node and is not allocated any power because it has either a high-quality direct link or a low-quality relay-to-destination link. Observe that the relay node considers *both* the quality of the direct links of the source nodes, and its own channel gain to the intended destinations, and will try to help the source nodes with low quality direct links, and high quality relay to destination links.

**Remark** 2: For RDF, the optimal power allocation tries to help the weak source nodes that it can efficiently assist, providing fairness among the source nodes. We note that, for low potential source nodes, the benefit provided by the relay node does not increase with increased relay power. Thus, an appropriate relay selection strategy for RDF relay networks should be to select the relay nodes that will provide both high quality source-to-relay and relay-to-destination links. When the relay power is scarce, the relay node will help only one source node that has the lowest $\frac{1+P_{si}\beta_i^2}{\gamma_i^2}$.

**Remark** 3: For the special case of the uplink of relay assisted TDMA network[2], the channel gains of all relay to destination links are equal, i.e., $\gamma_i = \gamma_k$ for the source nodes assisted by the $k$th relay node and we have (10) and (11) with $\gamma_i = \gamma_k \ \forall i$. An immediate corollary is that, in this case, the resulting high potential users all achieve identical capacities, i.e., we have $C_{i,RDF} = \frac{1}{2K} \log(\gamma_k^2/\mu_{k,RDF})$.

**Remark** 4: It is important to note that the optimum power allocation for the uplink of relay assisted TDMA network tries to equalize the individual capacities achieved by each source node, thus also increasing the symmetric capacity of the system. The source nodes that are not equalized in terms of individual capacities are the source nodes that either have a low quality source node to relay link (the set of low potential source nodes, and a subset of nonrelayed source nodes) or a very high quality direct link (nonrelayed source nodes).

### B. Nonregenerative Decode and Forward

When the relays assist the sources via NDF, we have the following theorem:

*Theorem 2:* The optimal power allocation for NDF relay networks also results in three source node sets, namely *high potential source nodes*, *low potential source nodes* and *nonrelayed* source nodes for each relay node. However, the allocated relay powers and the resulting capacities are different than that of RDF.

1) The optimum relay power dedicated to high potential source node $i$, and the achieved individual capacity of source node $i$, are

$$P_{ri} = (\frac{1}{\mu_{k,NDF}} - \frac{1}{\gamma_i^2})^+; \qquad C_{i,NDF} = \frac{1}{2K} \log(1 + P_{si}\beta_i^2) + \frac{1}{2K} \log(\gamma_i^2/\mu_{k,NDF}) \quad (17)$$

respectively, where $\mu_{k,NDF}$ is the water level for the $k$th NDF relay node that satisfies its power constraint.

2) The optimum relay power dedicated to low potential source node $i$, and the achieved individual capacity of source node $i$, are

$$P_{ri} = \frac{P_{si}(\alpha_i^2 - \beta_i^2)}{\gamma_i^2(1 + P_{si}\beta_i^2)}; \qquad C_{i,NDF} = \frac{1}{2K}\log(1 + P_{si}\alpha_i^2) \tag{18}$$

3) The nonrelayed source nodes are those source nodes for which either their direct links are better than their source-to-relay links, i.e., $\alpha_i^2 \leq \beta_i^2$, or their relay-to-destination links have low quality, i.e., $\frac{1}{\mu_{k,NDF}} \leq \frac{1}{\gamma_i^2}$.

*Proof:* The power allocation problem at the $k$th NDF relay node can be expressed as

$$\max_{\{P_{ri}\}_{i \in \hat{A}_k}} \sum_{i \in \hat{A}_k} \frac{1}{2K}\log(1 + P_{si}\beta_i^2) + \frac{1}{2K}log(1 + P_{ri}\gamma_i^2) \tag{19}$$

$$\text{s.t.} \sum_{i \in \hat{A}_k} P_{ri} \leq P_{Rk,total}; \qquad P_{ri} \geq 0, \quad \forall i \tag{20}$$

$$\frac{1}{2K}[log(1 + P_{si}\beta_i^2) + \log(1 + P_{ri}\gamma_i^2)] \leq \frac{1}{2K}\log(1 + P_{si}\alpha_i^2), \quad \forall i \tag{21}$$

The decodability constraint in (21) yields the upper bound

$$0 \leq P_{ri} \leq \frac{P_{si}(\alpha_i^2 - \beta_i^2)}{\gamma_i^2(1 + P_{si}\beta_i^2)} \tag{22}$$

The Lagrangian, $L(\{P_{ri}\}, \hat{\mu}_{k,NDF}, \{\rho_{i,NDF}\})$, is

$$\frac{1}{2K} \sum_{i \in \hat{A}_k} [log(1 + P_{si}\beta_i^2) + \log(1 + P_{ri}\gamma_i^2)] + \hat{\mu}_{k,NDF}(P_{Rk,total} - \sum_{i \in \hat{A}_k} P_{ri}) + \sum_{i \in \hat{A}_k} \rho_{i,NDF}(\frac{P_{si}(\alpha_i^2 - \beta_i^2)}{\gamma_i^2(1 + P_{si}\beta_i^2)} - P_{ri})$$

where $\hat{\mu}_{k,NDF}$ and $\rho_{i,NDF}$ are the Lagrange multipliers associated with the power constraint of the relay node $k$ and the upper bound for the relay power used for source node $i$, respectively. Once again, we have a convex program, and using KKT conditions, we arrive at the optimum relay power for source node $i$ as

$$P_{ri} = min((\frac{1}{\mu_{k,NDF}} - \frac{1}{\gamma_i^2})^+, \frac{P_{si}(\alpha_i^2 - \beta_i^2)}{\gamma_i^2(1 + P_{si}\beta_i^2)}) \tag{23}$$

where $\mu_{k,NDF} = 2K\hat{\mu}_{k,NDF}$. The source nodes for whom (21) is inactive, and $P_{ri} = \frac{1}{\mu_{k,NDF}} - \frac{1}{\gamma_i^2} > 0$ are high potential source nodes. When (21) is active, $P_{ri} = \frac{P_{si}(\alpha_i^2 - \beta_i^2)}{\gamma_i^2(1 + P_{si}\beta_i^2)}$, as the low potential source nodes. The source nodes that have $\frac{1}{\mu_{k,NDF}} \leq \frac{1}{\gamma_i^2}$ or $\alpha_i^2 \leq \beta_i^2$ are the nonrelayed source nodes. $\qquad\square$

**Remark** 5: Observe that the optimum power allocation for NDF relay networks tries to use the relay to destination channels as efficiently as it can without considering the direct links of the source nodes. The optimum solution is a modified water-filling solution with base levels $\frac{1}{\gamma_i^2}$, and upper levels $\frac{1}{\gamma_i^2} + \frac{P_{si}(\alpha_i^2 - \beta_i^2)}{\gamma_i^2(1 + P_{si}\beta_i^2)}$. The upper level is due to the decodability constraint of the NDF relay node. Such a power allocation scheme is demonstrated in Figure 2(b). In this example, source node 1 is a low potential source node and source nodes 2, 3 and 4 are high potential source nodes. Source node 5 is a nonrelayed source node since the relay node has very low channel gain to its destination. Similar to the RDF case, even if the total transmit power of relay nodes are increased, the low potential source nodes will not be able to achieve higher individual capacities. Thus, we can conclude that employing the appropriate relay selection strategy that provides high quality source node to relay, and relay to destination links, improves the performance of the NDF relay networks.

**Remark** 6: Similar to the RDF case in Remark 3, for the special case of uplink of relay assisted TDMA network, we can obtain the optimum power allocation for the NDF relay nodes by setting all $\gamma_i = \gamma_k$. In this case, we observe that the high potential source nodes become equal benefit source nodes. For this set of source nodes, the optimum power allocation does not depend on the transmission of the source nodes in the first time slots. The relay helps each user by increasing their rate by exactly $\frac{1}{2K} \log(1 + \frac{\gamma_k^2}{\mu_{k,NDF}})$. We also note that the nonrelayed user set is those nodes each of which has a better direct link than the relay link, $\alpha_i^2 \leq \beta_i^2$.

**Remark** 7: The optimum power allocation for the uplink of NDF relay assisted TDMA networks tries to equalize the benefits obtained by relay transmission for each source node. The optimum solution is a modified water-filling solution with upper bounds, $\frac{P_{si}(\alpha_i^2 - \beta_i^2)}{\gamma_k^2(1 + P_{si}\beta_i^2)}$ and *identical* base levels for each source node.

## IV. Relaying without Decode and Forward

### A. Amplify and Forward

When AF relay transmission is used, the individual capacity of source node $i$ is [5]

$$C_{i,AF} = \frac{1}{2K} \log(1 + P_{si}\beta_i^2 + \frac{P_{si}\alpha_i^2 P_{ri}\gamma_i^2}{P_{si}\alpha_i^2 + P_{ri}\gamma_i^2 + 1}) \qquad (24)$$

For AF relay assisted networks, we have the following theorem:

**Theorem 3:** The optimal power allocation for AF relay node results in nonzero power allocation for a subset of the source nodes assigned to the relay node. The optimum power allocated to assist source node $i$ is

$$P_{ri} = (\frac{-(\frac{a_i}{b_i} + 2) + \sqrt{(\frac{a_i}{b_i})^2 + \frac{4a_i}{\mu_{k,AF}}(1 + \frac{a_i}{b_i})}}{2(a_i + b_i)})^+ \tag{25}$$

where $a_i = \frac{P_{si}\alpha_i^2/(P_{si}\alpha_i^2 + 1)}{(1 + \beta_i^2 P_{si})/\gamma_i^2}$ and $b_i = \frac{\gamma_i^2}{P_{si}\alpha_i^2 + 1}$

while $\mu_{k,AF}$ is the water level for the $k$th AF relay node that satisfies $\sum_{i \in A_k} P_{ri} = P_{Rk,total}$.

*Proof:* The power allocation problem at the $k$th AF relay node can be expressed as

$$\max_{\{P_{ri}\}_{i \in A_k}} \sum_{i \in A_k} \frac{1}{2K} \log(1 + P_{si}\beta_i^2 + \frac{P_{si}\alpha_i^2 P_{ri}\gamma_i^2}{P_{si}\alpha_i^2 + P_{ri}\gamma_i^2 + 1}) \tag{26}$$

$$\text{s.t.} \sum_{i \in A_k} P_{ri} \le P_{Rk,total}; \qquad P_{ri} \ge 0, \quad \forall i \tag{27}$$

which, again is a convex program. The Lagrangian is

$$L(\{P_{ri}\}, \hat{\mu}_{k,AF}) = \frac{1}{2K} \log(1 + P_{si}\beta_i^2 + \frac{P_{si}\alpha_i^2 P_{ri}\gamma_i^2}{P_{si}\alpha_i^2 + P_{ri}\gamma_i^2 + 1}) + \hat{\mu}_{k,AF}(P_{Rk,total} - \sum_{i \in A_k} P_{ri})$$

where $\hat{\mu}_{k,AF}$ is the Lagrange multiplier associated with the total transmit power constraint of the relay node $k$. Simply taking the derivative with respect to $P_{ri}$ and equating it to zero, we arrive at the optimum relay power for source node $i$ in (25) where $\mu_{k,AF} = 2K\hat{\mu}_{k,AF}$.  □

**Remark** 8: The optimal power allocation for the AF relay nodes results in nonzero power allocation to the source nodes that satisfy $\mu_{k,AF} < a_i$. When the relay node is very close to a source node, then $a_i \approx \frac{\gamma_i^2}{P_{si}\beta_i^2 + 1}$ and $b_i \to 0$. This corresponds to the case when the source nodes' received SNR at the relay node are very high. The optimal power allocation in this case is identical to the optimal power allocation in RDF as expected. It is important to note that in AF, the individual capacities of the source nodes are not constrained by the capacity of the source node to relay channel. The upper bound for the individual capacity of source node $i$ is

$$C_{i,AF} \le \frac{1}{2K} \log(1 + P_{si}\beta_i^2 + P_{si}\alpha_i^2) \qquad \forall i \tag{28}$$

Thus, AF relaying may perform better than the DF relaying.

## B. Compress and Forward

In the case of CF relaying, when Gaussian codebooks are used, and the relay node compresses node $i$'s signal using Wyner-Ziv lossy source coding [18], the individual capacity of source node $i$ can be expressed as [17]

$$C_{i,CF} = \frac{1}{2K} \log(1 + P_{si}\beta_i^2 + \frac{P_{si}\alpha_i^2}{1 + \sigma_{Wi}^2}) \tag{29}$$

with

$$\sigma_{Wi}^2 = \frac{P_{si}(\alpha_i^2 + \beta_i^2) + 1}{P_{ri}\gamma_i^2(P_{si}\beta_i^2 + 1)} \tag{30}$$

For CF relay networks, we have the following theorem for optimum power allocation at each relay.

***Theorem 4:*** The CF relay with optimal power allocation assists a subset of the source nodes that are assigned to the relay node. The optimum power allocation for source node $i$ is

$$P_{ri} = (\frac{-(\frac{X_i}{Y_i} + 2) + \sqrt{(\frac{X_i}{Y_i})^2 + \frac{4X_i}{\mu_{k,CF}}(1 + \frac{X_i}{Y_i})}}{2(X_i + Y_i)})^+ \tag{31}$$

where $X_i = \frac{P_{si}\alpha_i^2\gamma_i^2}{(P_{si}\alpha_i^2 + P_{si}\beta_i^2 + 1)}$ and $Y_i = \frac{\gamma_i^2(P_{si}\beta_i^2 + 1)}{(P_{si}\alpha_i^2 + P_{si}\beta_i^2 + 1)}$ while $\mu_{k,CF}$ is the water level for the $k$th CF relay node that satisfies $\sum_{i \in A_k} P_{ri} = P_{Rk,total}$.

*Proof:* Proof follows identical steps to the proof of Theorem 3. □

**Remark** 9: Similar to AF case, the preassigned relay node $k$ allocates nonzero power to source node $i$ if $\mu_{k,CF} < X_i$. When $P_{Rk,total} \to \infty$, $\sigma_{Wi}^2 \to 0$, which yields the same asymptotic upper bound for the individual capacity of source node $i$ as in the AF case:

$$C_{i,CF} \leq \frac{1}{2K} \log(1 + P_{si}\beta_i^2 + P_{si}\alpha_i^2) \tag{32}$$

## V. Relay Strategy Selection

So far we investigated the power allocation problem for the relay nodes assisting all source nodes with the same relay transmission scheme. However, each relay transmission scheme has its own advantages and disadvantages, and one may overperform another in different scenarios. Thus, in principle, higher sum rates can be obtained if the relay has the flexibility to choose the appropriate relaying strategy for each source node it is assisting. In such a scenario, each source node will be relayed with the relaying strategy that will maximize its individual capacity resulting the individual capacity of source node $i$, i.e., $C_i = \max(C_{i,RDF}, C_{i,NDF}, C_{i,AF}, C_{i,CF})$. Formally, the power allocation problem with relaying strategy selection is

$$\max_{\{P_{ri}\}_{i=1,\cdots,K}} C_{sum} = \sum_{i=1}^{K} \max(C_{i,RDF}, C_{i,NDF}, C_{i,AF}, C_{i,CF}) \tag{33}$$

$$\text{s.t.} \quad \sum_{i \in A_k} P_{ri} \leq P_{Rk,total}; \quad P_{ri} \geq 0 \quad \forall i, k \tag{34}$$

Similar to the previous cases, we focus on the power allocation problem at each relay node. Using the inequality $\log(1 + X + Y) \leq \log(1 + X) + \log(1 + Y)$ for any $X, Y \geq 0$, it can be shown that $C_{i,RDF} \leq C_{i,NDF}$. Similarly, for all $P_{ri} \geq 0$, $C_{i,AF} \leq C_{i,CF}$. Thus, we can conclude that each source node should be relayed via either NDF or CF to maximize its individual capacity, $C_i = \max(C_{i,NDF}, C_{i,CF})$. The relaying strategy selection for each source node depends on $P_{ri}$ allocated for the transmission of the source node $i$'s signal. Thus, the power allocation policy at the relay nodes indirectly dictates the relaying strategy that the relay node should operate for each source node, and the total power of the relay node should be distributed appropriately considering these two relaying strategies.

It is readily seen that the optimization problem (33) is not a convex program. In this section, we seek a low complexity near-optimum algorithm to find the optimum power allocation with relaying strategy selection. For clarity of exposition, we denote the source node set assisted by the $k$th relay node in NDF relaying as $A_{k,NDF}$ and CF relaying as $A_{k,CF}$. We observe that once we fix $A_{k,NDF}$ and $A_{k,CF}$, the optimum power allocation problem becomes convex and the optimum solution is given by the following theorem.

*Theorem 5:* For a given relaying strategy selection for the $k$th relay node, $A_{k,NDF}$ and $A_{k,CF}$, the optimal relay power allocated to the source node $i$ relayed in NDF relaying and the optimal relay power allocated to the source node $j$ relayed in CF relaying are

$$P_{ri} = min\left(\left(\frac{1}{\mu_k} - \frac{1}{\gamma_i^2}\right)^+, \left(\frac{P_{si}(\alpha_i^2 - \beta_i^2)}{\gamma_i^2(1 + P_{si}\beta_i^2)}\right)^+\right) \tag{35}$$

$$P_{rj} = \left(\frac{-(\frac{X_j}{Y_j} + 2) + \sqrt{(\frac{X_j}{Y_j})^2 + \frac{4X_j}{\mu_k}(1 + \frac{X_j}{Y_j})}}{2(X_j + Y_j)}\right)^+ \tag{36}$$

respectively, where $\mu_k$ is the Lagrange multiplier associated with the total transmit power constraint of the relay node $k$.

*Proof:* The power allocation problem at the $k$th hybrid relay node with relaying strategy selection $A_{k,NDF}$ and $A_{k,CF}$ can be expressed as

$$\max_{\{P_{ri}\}_{i \in \hat{A}_k}} \quad \sum_{i \in A_{k,NDF}} C_{i,NDF} + \sum_{j \in A_{k,CF}} C_{j,CF} \tag{37}$$

$$\text{s.t.} \quad \sum_{i \in A_k} P_{ri} \leq P_{Rk,total}; \quad P_{ri} \geq 0 \quad \forall i, k \tag{38}$$

The Lagrangian, $L(\{P_{ri}\}, \hat{\mu}_k, \{\rho_i\})$, is

$$\frac{1}{2K} \sum_{i \in A_{k,NDF}} [log(1 + P_{si}\beta_i^2) + \log(1 + P_{ri}\gamma_i^2)] + \frac{1}{2K} \sum_{j \in A_{k,CF}} \log(1 + P_{sj}\beta_j^2 + \frac{P_{sj}\alpha_j^2}{1 + \sigma_{Wj}^2})$$

$$+ \hat{\mu}_k(P_{Rk,total} - \sum_{i \in \hat{A}_k} P_{ri}) + \sum_{i \in A_{k,NDF}} \rho_i(\frac{P_{si}(\alpha_i^2 - \beta_i^2)}{\gamma_i^2(1 + P_{si}\beta_i^2)} - P_{ri})$$

where $\hat{\mu}_k$ and $\rho_i$ are the Lagrange multipliers associated with the power constraint of the relay node $k$ and the upper bound for the relay power used for source node $i$ assisted in NDF relaying, respectively. Since the optimization problem is convex, using KKT conditions, we arrive at the optimum relay power for NDF relayed source node $i$ as in (35) and for CF relayed source node $j$ as in (36) with $\mu_k = 2K\hat{\mu}_k$. $\qquad \blacksquare$

Using Theorem 5, the jointly optimum power allocation and relaying strategy selection can be found comparing the performance of $2^K$ possible relaying strategy selection scenarios each of which would have a corresponding power allocation policy. However, the computational complexity of such an approach is too high. Thus, we seek a low complexity near-optimum algorithm. To that end, we first investigate the conditions under which one of NDF or CF would be preferred over the other.

When the direct link of a source node is better than the source to relay link, $\beta_i^2 \geq \alpha_i^2$, NDF relaying cannot improve the individual capacity of the source node resulting in $C_i = \max(C_{i,NDF}, C_{i,CF}) = C_{i,CF}$. Thus, the relay node should operate in CF mode for the source nodes with $\beta_i^2 \geq \alpha_i^2$. The set of such source nodes will be denoted as $A_{k,CF-strict}$ and the rest as $A'_{k,CF-strict}$ in the sequel. The relaying strategies that the relay node should employ for the source nodes $\beta_i^2 < \alpha_i^2$ are dependent on $\{P_{ri}\}$. $C_i$ versus $P_{ri}$ performance of a source node with $\beta_i^2 < \alpha_i^2$ is presented in Figure 3. For low $P_{ri}$ values NDF performs better than CF. At $P_{ri} = P_{ri-thre1}$, NDF relaying achieves its maximum capacity due to the decodability constraint of NDF relaying, i.e, $C_{i,NDF}(P_{ri-thre1}) = C_{upperDF}$ resulting $P_{ri-thre1} = \frac{P_{si}(\alpha_i^2 - \beta_i^2)}{\gamma_i^2(1 + P_{si}\beta_i^2)}$. Increasing $P_{ri}$ further without changing the relaying strategy for source node $i$ does not increase the individual capacity of source node $i$. Up to $P_{ri} = P_{ri-thre2}$, NDF still outperforms CF. For $P_{ri} = P_{ri-thre2}$, $C_{i,CF}(P_{ri-thre2}) = C_{i,NDF}(P_{ri-thre2}) = C_{upperDF}$ with $P_{ri-thre2} = \frac{(P_{si}(\alpha_i^2+\beta_i^2)+1)(\alpha_i^2-\beta_i^2)}{\gamma_i^2\beta_i^2(P_{si}\beta_i^2+1)}$. For $P_{ri} > P_{ri-thre2}$, CF performs better than NDF for source node $i$. Thus, for low relay power scenarios, relaying all of the source nodes in $A'_{k,CF-strict}$ with NDF and the rest with CF is a good relaying strategy, since it is likely that each source node will be allocated a relay power less than $P_{ri-thre2}$. Similarly, for high relay power, one may choose to relay all the source nodes with CF since it is likely that each source node will be allocated a relay power higher than $P_{ri-thre2}$. The global optimum power allocation with relaying strategy selection can be found via comparing the sum capacities of $2^{K-|A_{k,CF-strict}|}$ possible relaying strategy selection scenarios with their optimum power allocation.

Partitioning the source nodes into sets $A_{k,CF-strict}$ and $A'_{k,CF-strict}$ for CF and NDF relaying and finding the optimum power allocation for such a relaying strategy selection can be a strong candidate for the jointly optimum power allocation and relaying strategy selection: If the optimum power allocation for such a partition results in all the NDF decodability constraints of the source nodes in $A'_{k,CF-strict}$ being non-active, then the global optimum power allocation with relaying strategy selection is found. This is due to the fact that this solution is also the solution of the optimization problem when we relax the NDF decodability constraints of the original power allocation problem in (33) which provides an upper bound for the original problem. Such an approach is especially useful for low relay power scenarios where the probability of $A'_{k,CF-strict}$ being the optimum $A_{k,NDF}$ is very high. If the NDF decodability constraints become active for some of the source nodes with such a relaying strategy, partitioning the source nodes into $A_{k,CF-strict}$ and $A'_{k,CF-strict}$ for CF and NDF relaying, may not be the optimum relaying strategy. However, investigating the optimal power allocation policy for such a partition provides insight towards the optimum relaying strategy. It is important to note that more relay power should be dedicated to a source node's transmission if it is relayed with CF relaying strategy. Thus, if a source node switches from NDF to CF, then less relay power will be allocated to the remaining source nodes. Since some of the source nodes have to switch from NDF mode to CF for the optimum relaying strategy, the source nodes that have non-active decodability constraints with the optimum power allocation of $A_{k,DF} = A'_{k,CF-strict}$ and $A_{k,DF} = A_{k,CF-strict}$ will be allocated less relay power for the optimum relaying strategy, and the decodability constraints of these source nodes will still be non-active for the optimum relaying strategy. These source nodes should be relayed with NDF also for the optimum relaying strategy selection, and will be denoted as $A_{k,NDF-strict}$ in the sequel. Using this observation, we obtain the optimum relaying strategies for the source nodes in $A_{k,CF-strict}$ and $A_{k,DF-strict}$. Furthermore, we propose the following strategy to enable switching from NDF to CF for the source nodes that are neither in $A_{k,CF-strict}$ nor $A_{k,NDF-strict}$. Noting that the cost of switching from NDF to CF of a source node in terms of power consumption is $P_{ri-thre2} - P_{ri-thre1}$, we choose the source node that has the lowest $P_{ri-thre2} - P_{ri-thre1}$ for switching from NDF to CF, and check if the optimum power allocation for such a switch in the relaying strategy results in increased sum capacity. We continue to switch the source nodes to CF until switching a source node from NDF to CF does not improve the sum capacity or all source nodes are switched to CF except the source nodes in $A_{k,NDF-strict}$. The outline of the near-optimum relaying strategy selection algorithm (NORSS) is summarized in Table I.

## VI. NUMERICAL RESULTS

In this section, we present numerical results to demonstrate the performance of optimum power allocation and relaying strategy selection for a relay assisted F/TDMA ad hoc network. For numerical results, we consider an ad hoc F/TDMA multiuser relay network with 4 source nodes and one relay node that serves all in one of the RDF, NDF, AF, CF relaying strategies for each source node. The link SNRs of the source nodes used throughout the simulations are $\{(P_{si}\beta_i^2, P_{si}\alpha_i^2, \gamma_i^2)\}_{i=1}^4 = \{(12.25, 19.51, 11.84), (7.03, 16.45, 7.03), (9.03, 11.84, 18.06), (8.06, 9.03, 16.45)\}$ dB. We investigate the individual capacities and the sum capacities resulting from the proposed power allocation schemes and relaying strategy selection algorithm (NORSS) for a range of relay power constraints.

Figures 4, 5, 6, 7 and 8 show the performance of the individual capacities of the source nodes achieved by pure RDF, pure NDF, pure AF, pure CF and hybrid relaying (NORSS) with optimum power allocation, respectively. We observe that the individual capacities are improved as the relay power is increased up to a threshold for each source node. In the RDF case, when the relay node has relatively low power, the relay node helps only the third and fourth source nodes that have relatively high $\frac{\gamma_i^2}{1+P_{si}\beta_i^2}$, since the rest of the source nodes have higher direct links or the relay has low quality links to the destinations of these source nodes. As the available power at the relay increases, the third and fourth source nodes' potential are reached, and the relay node starts to help the rest of the source nodes. We also observe that after a threshold, increasing the relay power does not help, since all source nodes already achieve the maximum single-source node capacities. In the NDF case, we again observe that the sum capacity is improved as the relay power is increased up to a threshold. Since NDF performs better than RDF, this threshold is much lower than the threshold in the RDF case. That is, for the maximum sum capacity, NDF requires less power at the relay node as compared to RDF. For NDF, we observe that the relay tries to use the relay to destination channels as efficiently as it can, without considering the performance of the direct links. However, the benefit that can be provided by the relay node is limited by the quality of the source node to relay link. In Figure 6, we observe that the benefit obtained by the AF relay nodes converges to its maximum point gradually for each source node. Similar behavior is observed for the CF relay transmission in Figure 7. We also observe that in the AF and CF relaying, both individual capacities and the resulting sum capacities may be higher than the capacities that result from operating in a DF relaying. This is due to the fact that DF relaying has the decodability constraints in the source node to relay links whereas the AF and CF do not. Figure 8 shows the performance of the individual capacities of the source nodes achieved by NORSS with optimum power allocation. We observe that, in the low relay power case, source nodes 3 and 4 are assisted in NDF relaying strategy whereas no relay power is dedicated to the 1st and 2nd source nodes. This is due to the fact that the source nodes 3 and 4 have much better relay to destination links than the source nodes 1 and 2. Thus, it is not efficient to allocate power to the source nodes 1 and 2 in low relay power case. As the available power at the relay node increases, the relay node starts to help the 1st and 2nd source nodes. We observe that after some threshold, the potentials of the source nodes for NDF relaying strategy have been reached, and the relay node switches to CF relaying strategy after enough relay power becomes available for each source node. As expected in low relay power cases, NDF relaying strategy is preferred whereas in high relay power cases, the relay node switches to CF relaying strategy to provide higher capacities. Figure 9 demonstrates the sum capacities resulting from NORSS, pure RDF, NDF, AF and CF relaying strategies with optimum power allocation. As expected NORSS uses the advantages of both NDF and CF relaying, and performs better than pure RDF, NDF, AF or CF relaying. Note that for low relay power cases, hybrid relaying favors NDF, and for high relay power cases, hybrid relaying uses CF for all source nodes to obtain higher sum capacities. Observe also that NORSS finds the optimum relaying strategy for each source node.

## VII. CONCLUSION

In this paper, we have considered a two-hop multiple source-destination F/TDMA wireless network where intermediate nodes relay the information of source nodes. We have solved the problem of optimally allocating the power of each relay node between the source nodes' transmissions it is assisting for different relay transmission schemes. We first investigated the problem for the relay nodes where each source node is assisted via the same relaying strategy. We have observed that the optimum power allocation for RDF relay nodes helps the source nodes that have low quality direct links and have destinations near to the relay first, and tries to improve the individual capacities of the weak source nodes. The optimum power allocation in NDF relay networks tries to use the relay to destination channels as efficiently as it can. We also observe that the AF and CF relay nodes provide higher sum capacities than the DF relay nodes with high relay powers due to the decodability constraints of DF relaying. Motivated by higher sum capacities, we have then investigated the power allocation problem with relaying strategy selection where the hybrid relay nodes chooses the best relaying strategy for each source node, and proposed a near-optimum relaying strategy selection algorithm. We have observed that hybrid relaying with the near-optimum relaying strategy selection algorithm and optimum power allocation performs better than pure RDF, NDF, AF or CF relaying with optimum power allocation.

In this paper, we aim to establish performance limits of two hop networks and therefore the results obtained in this paper come with the usual information theoretic disclaimers. We also note that the formulation assumes pre-assigned relay nodes, i.e., fixed routing decision. An interesting extension is the problem of jointly optimum relay node selection for each source node along with the relaying strategy and the power allocation.

## REFERENCES

[1] M. Haenggi. Analysis and design of diversity schemes for ad hoc wireless networks. *IEEE Journal on Selected Areas in Communications*, 23(1):19 − 27, January 2005.

[2] O. Leveque and I. E. Telatar. Information-theoretic upper bounds on the capacity of large extended ad hoc wireless networks. *IEEE Transactions on Information Theory*, 51(3):858 − 865, March 2005.

[3] E. van der Meulen. A survey of multi-way channels in information theory: 1961-1976. *IEEE Transactions on Information Theory*, 23(1):1 − 37, January 1977.

[4] T. M. Cover and A. A. El Gamal. Capacity theorems for the relay channel. *IEEE Transactions on Information Theory*, 25(5):572 − 584, September 1979.

[5] J. N. Laneman, D. N. C. Tse, and G. W. Wornell. Cooperative diversity in wireless networks: Efficient protocols and outage behavior. *IEEE Transactions on Information Theory*, 50(12):3062 − 3080, December 2004.

[6] A. Sendonaris, E. Erkip, and B. Aazhang. User cooperation diversity, part I: System description, part II: Implementation, aspects and performance analysis. *IEEE Transactions on Communications*, 51(11):1927 – 1938, November 2003.

[7] I. Maric and R. D. Yates. Bandwidth and power allocation for cooperative strategies in Gaussian relay networks. In *38th Asilomar Conference on Signals, Systems and Computers*, November 2004.

[8] D. R. Brown III. Resource allocation for cooperative transmission in wireless networks. In *38th Asilomar Conference on Signals, Systems and Computers*, November 2004.

[9] M. O. Hasna and M. S. Alouini. Optimal power allocation for relayed transmissions over Rayleigh-fading channels. *IEEE Transactions on Wireless Communications*, 3(6):1999 – 2004, November 2004.

[10] J. Boyer, D. D. Falconer, and H. Yanikomeroglu. Multihop diversity in wireless relaying channels. *IEEE Transactions on Communications*, 52(10):1820 – 1830, October 2004.

[11] A. Host-Madsen and J. Zhang. Capacity bounds and power allocation in wireless relay channel. *IEEE Transactions on Information Theory*, 51(6):2020 – 2040, June 2005.

[12] I. Maric and R. D. Yates. Cooperative multicast for maximum network lifetime. *IEEE Journal on Selected Areas in Communications*, 23(1):127 – 135, January 2005.

[13] A. Scaglione and Hong Yao-Win. Opportunistic large arrays: cooperative transmission in wireless multihop ad hoc networks to reach far distances. *IEEE Transactions on Signal Processing*, 51(8):2082 – 2092, August 2003.

[14] G. Kramer and A. J. van Wijngaarden. On the white Gaussian multiple-access relay channel. In *IEEE International Symposium on Information Theory*, June 2000.

[15] P. Gupta and P. R. Kumar. Towards an information theory of large networks: an achievable rate region. *IEEE Transactions on Information Theory*, 49(8):1877 – 1894, August 2003.

[16] G. Kramer, M. Gastpar, and P. Gupta. Cooperative strategies and capacity theorems for relay networks. *IEEE Transactions on Information Theory*, 51(9):3037 – 3063, September 2005.

[17] L. Sankaranarayanan, G. Kramer, and N. B. Mandayam. Capacity theorems for the multiple-access relay channel. In *42nd Annual Allerton Conference on Communications, Control and Computers*, September 2004.

[18] A. Wyner and J. Ziv. The rate-distortion function for source coding with side information at the decoder. *IEEE Transactions on Information Theory*, 22(1):1 – 10, January 1976.

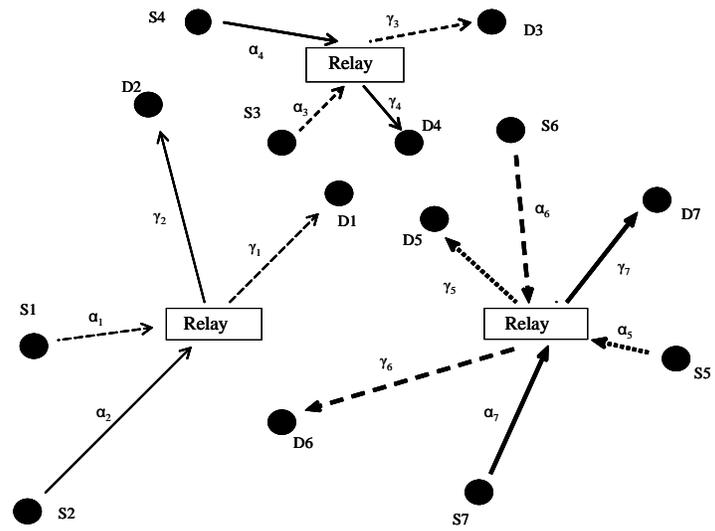

Fig. 1.   System Model

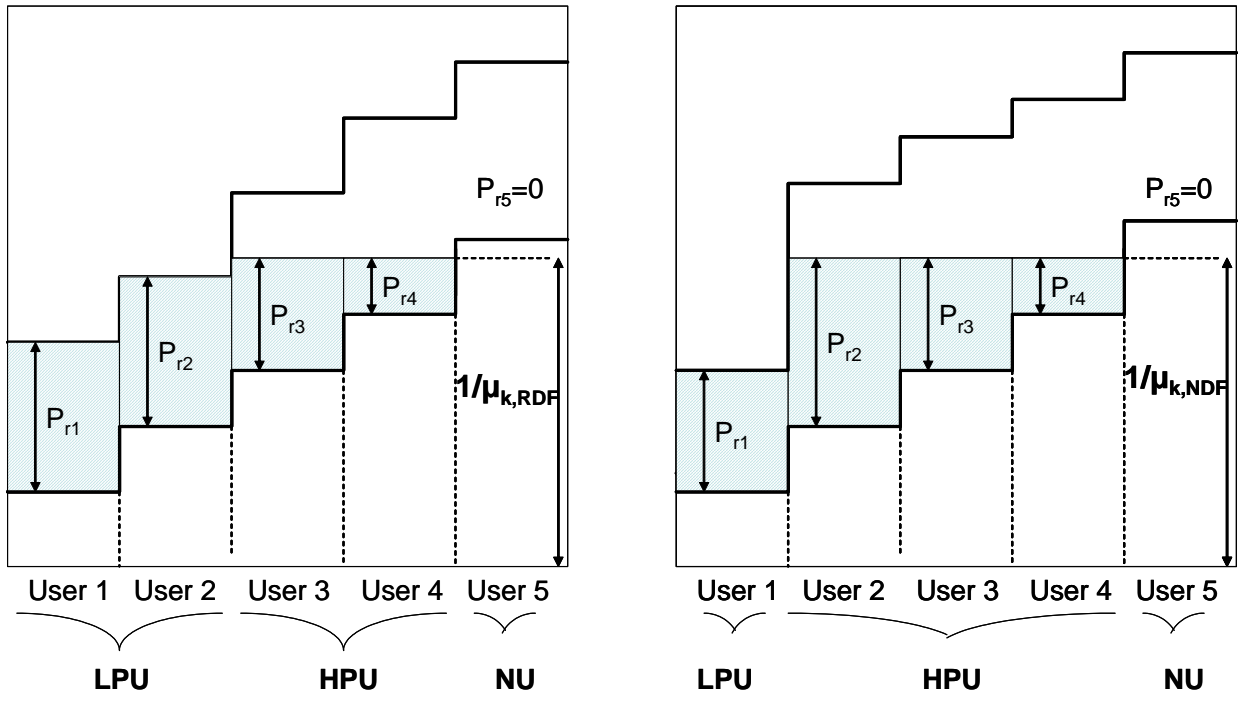

(a) RDF Relaying.                              (b) NDF Relaying.

Fig. 2.   Modified water-filling solution for Decode-and-Forward relaying

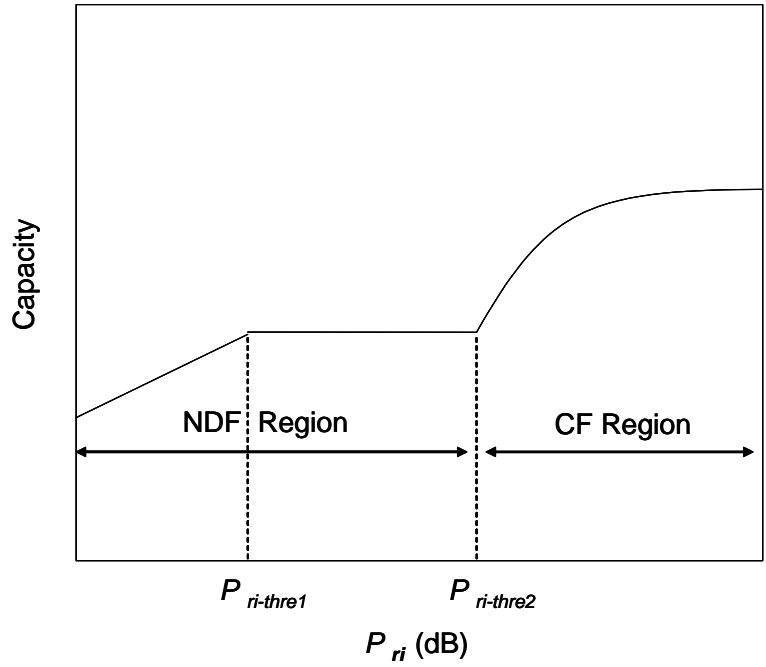

Fig. 3. Relay Power vs Capacity for Hybrid relaying

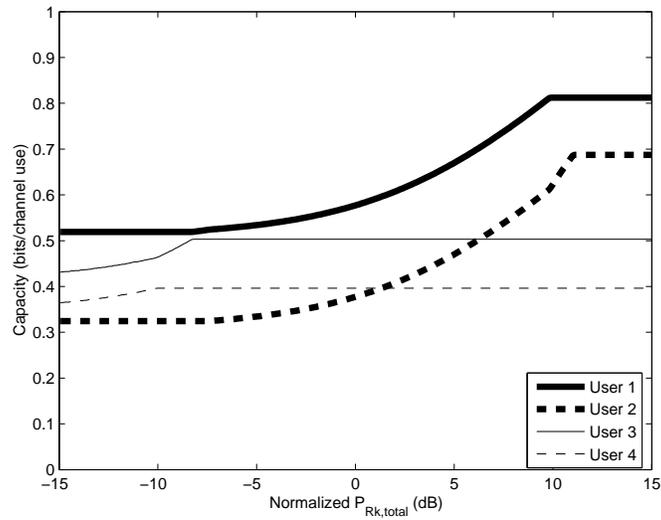

Fig. 4. RDF relay networks

TABLE I

NEAR OPTIMUM RELAYING STRATEGY SELECTION

Step-1: Form $A_{k,CF-strict}$ set
    $A_{k,CF-strict} = \emptyset$
    For $i \in A_k$
        if $\beta_i^2 \geq \alpha_i^2$
            $A_{k,CF-strict} = A_{k,CF-strict} \cup \{i\}$
        end
    end

Step-2: Form $A_{k,DF-strict}$ set
    $A_{k,CF} = A_{k,CF-strict}$
    $A_{k,DF} = A_k - A_{k,CF-strict}$
    Find the optimal power allocation with the defined $A_{k,CF}$ and $A_{k,DF}$ sets
    $\{P_{ri}^*\} = \underset{\{P_{ri}\}_{i \in A_k}}{\arg \max} \ C_{sum}(A_{k,CF}, A_{k,DF})$
    $SumCap = \underset{\{P_{ri}\}_{i \in A_k}}{\max} \ C_{sum}(A_{k,CF}, A_{k,DF})$
    Check the decodability constraints of the source nodes in $A_{k,DF}$
    $A_{k,DF-strict} = \emptyset$
    For $i \in A_{k,DF}$
        If (Decodability Constraint of source node $i$ is passive)
            $A_{k,DF-strict} = A_{k,DF-strict} \cup \{i\}$
        end
    end

Step-3: For the source nodes in $A_{k,DF} - A_{k,DF-strict}$ , check sequentially if switching from DF to CF improves the sum capacity
    Do While $(A_{k,DF} - A_{k,DF-strict} \neq \emptyset)$
        Find the source node with the lowest power cost for DF to CF switch
        $j = \underset{i \in (A_{k,DF}-A_{k,DF-strict})}{\arg \min} \ P_{ri-thre2} - P_{ri-thre1}$
        Form the candidate $A_{k,DF}$ and $A_{k,CF}$ sets with the switch of source node $j$ from DF to CF
        $A_{k,DF-cand} = A_{k,DF} - \{j\}$
        $A_{k,CF-cand} = A_{k,CF} \cup \{j\}$
        Find the maximum sum capacity with the optimal power allocation of $A_{k,DF-cand}$ and $A_{k,CF-cand}$ sets
        $SumCap - cand = \underset{\{P_{ri}\}_{i \in A_k}}{\max} \ C_{sum}(A_{k,CF-cand}, A_{k,DF-cand})$
        Check if the switch of source node $j$ from DF to CF improves the sum capacity or not and update $A_{k,DF}$, $A_{k,CF}$ and $A_{k,DF-strict}$ sets.
        If $(SumCap - cand > SumCap)$
            $A_{k,DF} = A_{k,DF-cand}$
            $A_{k,CF} = A_{k,CF-cand}$
            $SumCap = SumCap - cand$
          else
            $A_{k,DF-strict} = A_{k,DF-strict} \cup \{j\}$
        end
    end
end

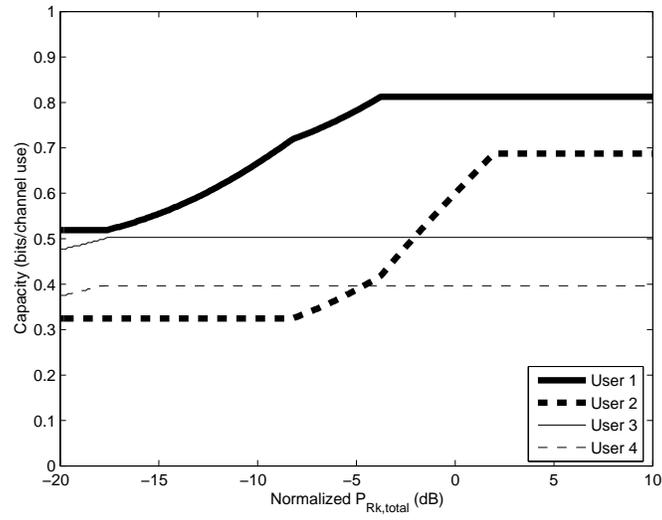

Fig. 5. NDF relay networks

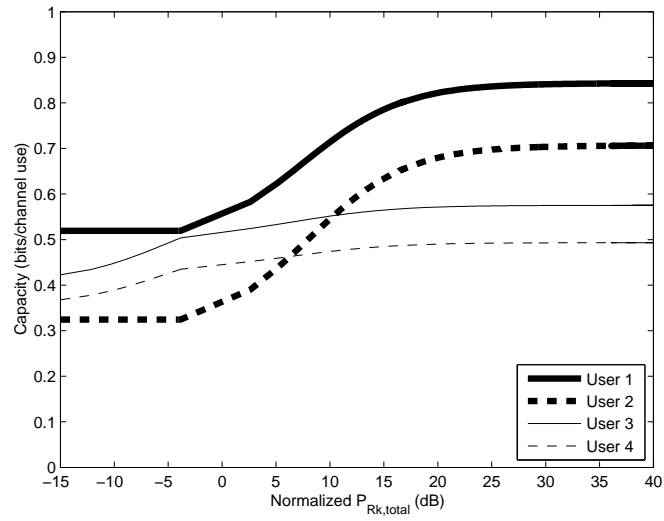

Fig. 6. AF relay networks

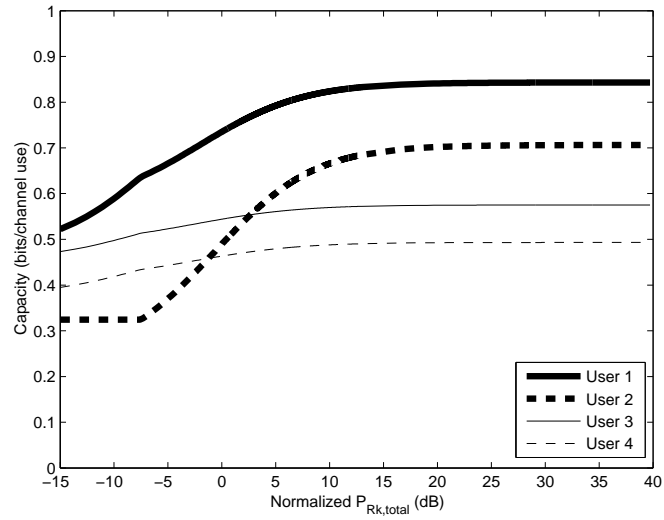

Fig. 7. CF relay networks

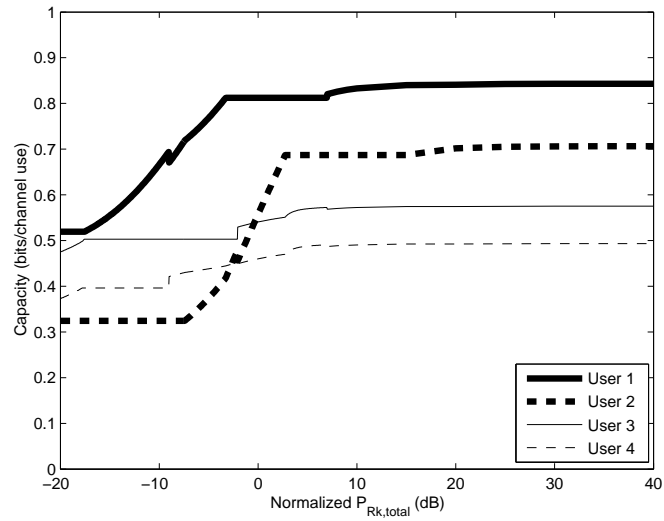

Fig. 8. Hybrid relay networks

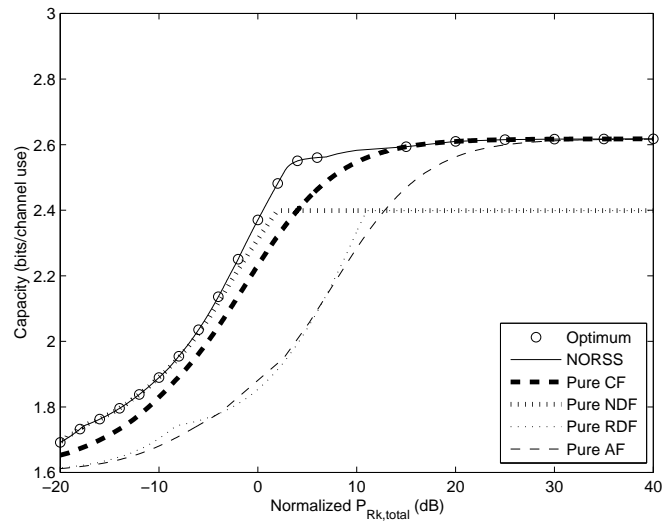

Fig. 9. Comparison of relaying strategies